\def \cm{~\rm{cm}}
\def \s{~\rm{s}}

\def \km{~\rm{km}}

\def \g{~\rm{g}}

\def \erg{~\rm{erg}}

\documentclass{iaus}
\usepackage{graphicx}
\begin{document}

\title[Exploding SNe with jets] 
{Exploding SNe with jets: time-scales}

\author[Oded Papish \& Noam Soker]   
{Oded Papish$^1$
 \and Noam Soker$^1$}

\affiliation{$^1$Dept. of Physics, Technion, Haifa 32000, Israel;
papish@physics.technion.ac.il; soker@physics.technion.ac.il.}

\pubyear{2012}
\volume{279}  
\pagerange{xxx--yyy}
\setcounter{page}{1}
\jname{Death of Massive Stars: Supernovae and Gamma-Ray Bursts}
\editors{P. Roming, N. Kawai \& E. Pian, eds.}

\maketitle

\begin{abstract}
We perform hydrodynamical simulations of core collapse supernovae (CCSNe) with a cylindrically-symmetrical numerical code (FLASH) 
to study the inflation of bubbles and the initiation of the explosion  within the frame of the jittering-jets
model. We study the typical time-scale of the model and compare it to the typical time-scale of the delayed neutrino mechanism.
Our analysis shows that the explosion energy of the delayed neutrino mechanism is an order of magnitude less than the required $10^{51} \erg$.

\end{abstract}

\section{The Jittering-Jets Model}
Accretion-outflow systems are commonly observed in astrophysics when compact objects accrete mass via an accretion disk that launches jets.
This is the basic engine assumed in jet-based CCSN models (\cite[Papish \& Soker 2011, Lazzati et al. 2011]{Papish2011, Lazzati2011}).
Our \textit{Jittering-Jet Model} for CCSN explosions is based on the following ingredients.
(1) We do not try to revive the stalled shock. To the contrary, our model requires the material near the stalled-shock
to fall inward and form an accretion disk around the newly born neutron star (NS) or black hole (BH).
(2) We conjecture that due to stochastic processes and the stationary
accretion shock instability (SASI; e.g., \cite[Blondin \& Shaw 2007, Fern{\'a}ndez 2010]{Blondin2007,Fernandez2010}) segments of the post-shock accreted gas possess local angular momentum (see also \cite[Foglizzo et al. (2012)]{Foglizzo2012} for an experimental demonstration).
(3) We assume that the accretion disk launches two opposite jets. Due to the rapid change in the disk's axis, the
jets can be intermittent and their direction rapidly varying. These are termed jittering jets.
(4) The jets penetrate the infalling gas up to a distance of few$\times 1000 \km$, i.e.,
beyond the stalled-shock.
The jets deposit their energy inside the star via shock waves, and form hot bubbles.
(5) The jets are launched only in the last phase of accretion onto the NS.

We perform 2.5D numerical simulations using the FLASH code (\cite[Fryxell et al. 2000]{Fryxell2000}).
We use 2D cylindrical coordinates on a grid of size $1.5 \times 10^9 \cm$
in each direction. We use 10 mesh grid refinements which gives us a resolution of $3 \km$ at the inner boundary of the grid of $r=70 \km$.
For the initial conditions of the ambient gas in the core we used the $15 M_\odot$ model of
\cite[Liebend{\"o}rfer et al. (2005)]{Liebend2005} who made a 1D simulations of the core bounce.
We start the simulation by launching jets at $0.25 \s$ after bounce.
We inject a jet (only one jet as only one half of space is simulated) at $75 \km$ from the center with a full opening angle of $10$ degrees.
To simulate the jittering effect we inject the jet for
a time interval of $\Delta t_j = 0.05 \s$ with a constant angle $\theta_n$ relative to the $z$ axis.
After this time interval, we stop the jets for a period of  $\Delta t_p = 0.05 \s$.
We then continue with a jet at a different angle of $\theta_{n+1}$, for the same $\Delta t_j = 0.05 \s$.
We repeat this process for several times.
The velocity of the jet is taken to be $10^{10} \cm \s^{-1}$ with with a mass outflow rate of 
$\dot M_{2j} = 4 \times 10^{31} \g \s^{-1}$.
The total energy carried by the two opposite jets combined over all episodes is
$E_{2j} = 2 \times 10^{51} \erg$. 
Our preliminary results are presented in Fig. 1.
\begin{figure}
  \centering
    \includegraphics[width=0.45\textwidth]{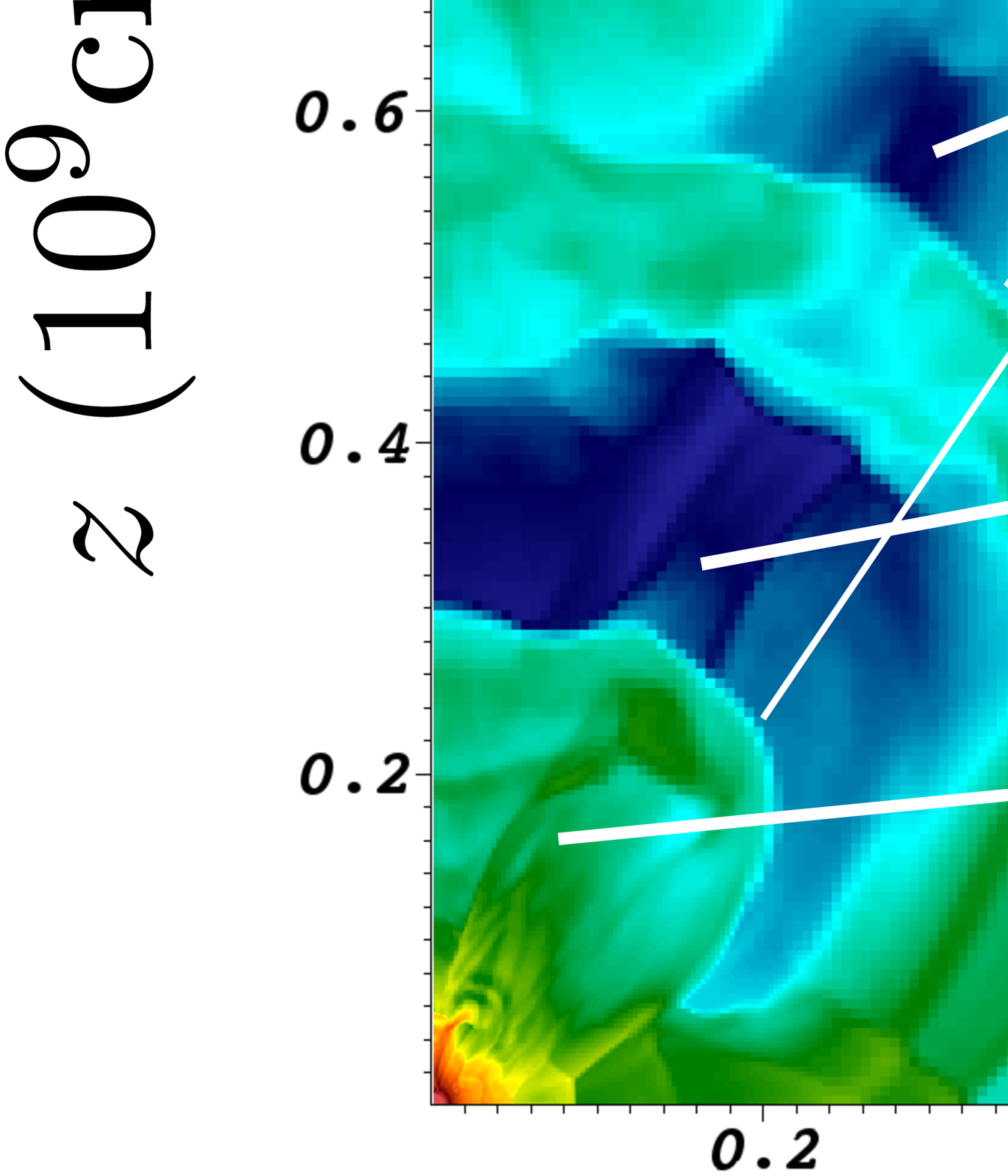}
        \includegraphics[width=0.45\textwidth]{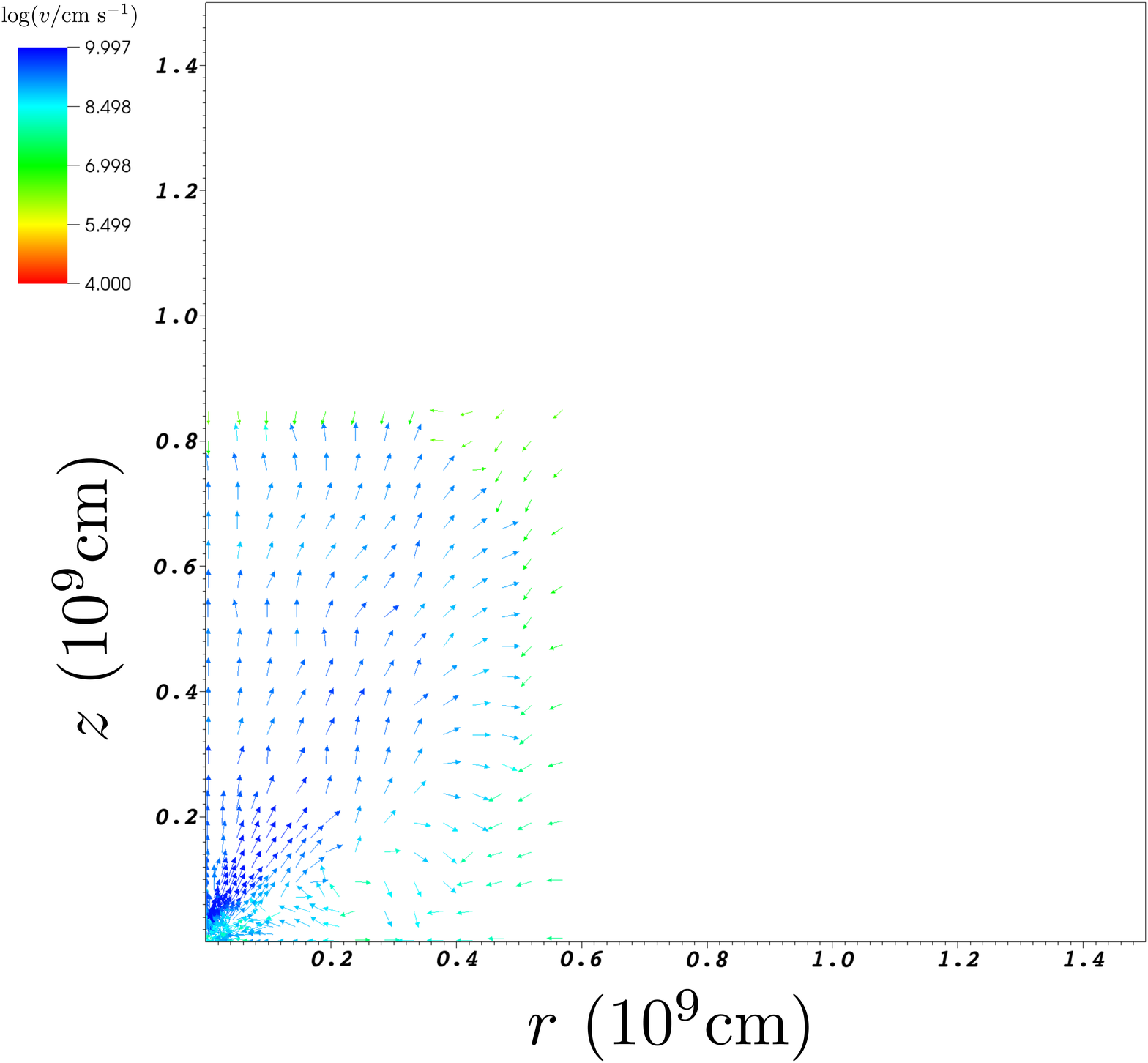}

      \caption{Time $t = 0.61 \s$ after bounce with the first jets episode starts at $t = 0.25 \s$ after bounce. $4$ episodes of jets were launched up to this time, in directions $\theta=10^\circ, 30^\circ, 50^\circ,$ and $35^\circ$. The first two episodes are already mixed together; the 4th jet's episode was turned off 0.01 seconds ago. Note that accretion from the equatorial plane (the horizontal plane in the figure) continues alongside with the outflow induced by the jets.
       The left plot shows the density, the right plot shows the log of the velocity.
      All values are in cgs units. The size of the grid is $1.5\times 10^9 \cm$ in each direction.}
      \label{fig:bubble}
\end{figure} 
\section{Time-scale Considerations}
The typical parameters of the model are the following.
The active phase of the jets lasts for a typical time equals to few times the free fall time from the region where
the baryonic mass is $\sim 1.5-1.6 M_\odot$. This radius is $\sim 3000 \km$ and the free fall time is $0.4 \s$.
The jets deposit their energy in a typical radius of $r_{d} \simeq 10^4 \km$. The average power of the jets
is taken to be $P_{2j} \simeq 10^{51} \erg \s^{-1}$. The jets are active for $t_j \simeq 1-2 \s$.
This is also the dynamical time at $r_d$
\begin{equation}
t_d \equiv \frac{r_d}{v_{\rm esc}(r_d)} = 1.6 \left(\frac{r_d}{10^4 \km} \right)^{3/2} \s,
\label{eq1}
\end{equation}
where $v_{\rm esc}$ is the escape velocity at $r_d$, and where the central mass is taken to be $1.5 M_\odot$.
The jets' power and the interaction time at $r_d \sim 10^4 \km$ are consistent with the energy required to explode the star.
From this time on, the hot bubbles (which might merge to one bubble) explode the star.

We find that the same consideration of interaction time and explosion power is problematic for neutrino driven CCSN explosion models.
In recent numerical results {(e.g. \cite[Brandt et al. 2011, Hanke et al. 2011, Kuroda et al. 2012]{Brandt2011, Hanke2011, Kuroda2012}) }the mechanical energy achieved
by neutrino driven models is still significantly short of the desired
$\sim 10^{51} \erg$ required to explode the star.
In the recent simulations of \cite[Mueller et al. (2012)]{Mueller2012} the sum of the kinetic, thermal, and nuclear energy of the expanding gas in the core
is a factor of $\sim 4$ smaller than the observed energy of CCSNe.
{We note that in \cite[Nordhaus et al. (2011) and Scheck et al. (2006)]{Nordhaus2011, Scheck2006} the explosion is achieved mainly by a continues wind. Here we refer to the delayed neutrino mechanism models where the energy of the
wind is negligible.}

Let us apply the interaction time considerations to the delayed neutrino mechanism.
The gain region, where neutrino heating is efficient, occurs in the region $r \simeq 200-700 \km$ (\cite[Janka 2001]{Janka2001}).
If energy becomes significant the gas will be accelerated and escape within a time of $\sim r/[0.5 v_{\rm esc} (r)]$. From \cite[Janka (2001)]{Janka2001} we find the neutrino ``optical depth'' from $r$ to infinity to be $\tau \sim (r/100 \km)^{-3}$.
The typical electron (and positron) neutrino luminosity is $L_{\nu} \simeq 5 \times 10^{52} \erg \s^{-1}$ (\cite[Mueller et al. 2012]{Mueller2012}).
Over all, if the interaction occurs near a radius $r$ in the gain region, the energy that can be acquired by the expanding gas is
\begin{equation}
E_{\rm shell} \simeq \frac {2 r}{v_{\rm esc} (r)} \tau L_{\nu}
\simeq 10^{50} \left( \frac{r}{200 \km} \right)^{-1.5} \left( \frac{L_{\nu}}{5 \times 10^{52} \erg \s^{-1}} \right) \erg.
\label{eq2}
\end{equation}
We claim, therefore, that the total energy that can be used to revive the shock is limited to
a typical value of $\sim 10^{50} \erg$.
This is along the recent results of numerical simulations of the delayed neutrino mechanism cited above.

{We conclude that the delayed neutrino explosion mechanism, where the
explosion is due to neutrino heating in the gain region, as proposed by
Wilson (Bethe \& Wilson 1985), cannot work. 
It might lead to the reviving of the stalled shock under some circumstances,
but it cannot lead to an explosion with an energy of $10^{51} \erg$"}

Our basic conclusion is that { if no ingredient is added to the neutrino delayed mechanism, it falls short by an order
of magnitude from the required energy to explode the star.
Such an ingredient can be a strong wind, such
as was applied by artificial energy deposition (\cite[Nordhaus et al. 2011]{Nordhaus2011}).
In their 2.5D simulations \cite[Scheck et al. (2006)]{Scheck2006} achieved explosion that was mainly driven by a
continues wind. The problem we see with winds is that they are less efficient than jets. 
Indeed, in order to obtain an explosion the winds in the simulations of \cite[Scheck et al. (2006)]{Scheck2006} had to be massive. For that, in cases where they obtained energetic enough explosions the final mass of the NS was low  $(M_{\rm NS} <1.3 M_\odot)$.
The problem we find is that such a wind must be active while accretion takes place; the accretion is required for the energy source.}

An inflow-outflow situation naturally occurs with jets launched by accretion disks.
For that, we propose the jittering-jet mechanism.
Namely, we require the accretion process to continue for $\sim 1 \s$.
In our model there is no need to revive the accretion shock.
{
\acknowledgements
We thank Thomas Janka, Kei Kotake, and Jason Nordhaus for useful comments.
This research was supported by the Asher Fund for Space
Research at the Technion, and by the Israel Science Foundation.
}

\end{document}